\documentclass[12pt]{article}
\usepackage{eepic, epic, epsfig, amsmath}

\newcommand{\sect}{\section}

\title{Quantum Field Theory of Forward Rates with Stochastic Volatiltiy}

\author{Belal E. Baaquie\\
{\small Department of Physics},\\
{\small National University of Singapore, Kent Ridge Road,}\\ 
{\small Singapore 091174, Republic of Singapore}
\\{\small e-mail:uspbeb@nus.edu.sg}}
\begin{document}

\maketitle

\begin{abstract}

In a recent formulation of a quantum field theory of forward 
rates, the volatility of the forward rates was taken to be deterministic. The field theory of the forward rates is generalized to the case of stochastic volatility. Two cases are analyzed, firstly when volatility is taken to be a function of the forward rates, and secondly when volatility is taken to be an independent quantum field. Since volatiltiy is a positive valued quantum field, the full theory turns out to be an interacting nonlinear quantum field theory in two dimensions. The state space and Hamiltonian for the interacting theory are obtained, and shown to have a nontrivial structure due to the manifold moving with a constant velocity. The no arbitrage condition is reformulated in terms of the Hamiltonian of the system, and then exactly solved for the nonlinear interacting case.

\end{abstract}

\newpage
\sect{Introduction}

Forward rates are essential to the debt market, and have wide-ranging applications in finance. The most widely used model of the forward rates is the the Heath-Jarrow-Morton (HJM) \cite{hjm}. There are a number of ways that the HJM-model can be generalized. In references \cite {s3} and \cite {s4} a correlation between forward rates with different maturities was introduced, and in \cite{s1} and \cite{s2} the forward rate was modeled as a stochastic string.  

The application of techniques of physics to finance \cite{bs}, \cite{jp} have proved to be a fruitful field; in particular path integral techniques \cite{b} have been applied to various problems in finance. In \cite{beb2} path integral techinques were applied to study the case of a security with stochastic volatility. In \cite{beb1} the HJM-model was generalized by treating the forward rates as a quantum field; empirical studies in \cite{b2} show that the field theoretic  model for the forward rates proposed in \cite{beb1} fits market data fairly well. 

The volatility of the forward rates is a central measure of the degree to which the forward rates fluctuate. In the model studied in \cite{beb1}, the volatiltity of the forward rates was taken to be deterministic. The question naturally arises as to whether the volatility itself should considered to be a randomly fluctuating quantity. The volatility of volatility is an accurate measure of the degree to which volatility is random. Market data for the Eurodollar futures provides a fairly accurate estimate of the forward rates for the US dollar, and also yields its volatility of volatility of the forward rates.

\begin{figure}[h]
 \begin{center}
   \epsfig{file=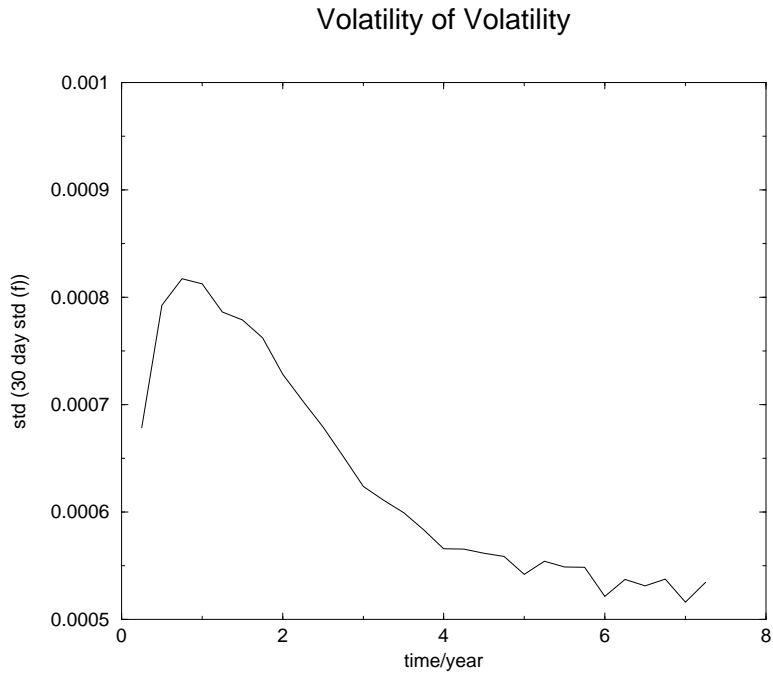, width=9cm, angle=-90}
     \caption{The Volatility of the Volatility of the Forward Rates}
   \label{volvol}
  \end{center}
\end{figure}

Eurodollar futures data given in Figure \ref{volvol} plots the $30$ days moving average of the volatility of volatility for the forward rates, and shows that it contributes about $0.0006-0.0007$ per year to the forward rates, which are in the neighbourhood of $0.05-0.06$ per year. The fluctuations in the volatility of the forward rates are about $10\%$ of the forward rates, and hence significant.

We conclude from the data that the volatility of the forward rates needs to be treated as a fluctuating quantum (stochastic) field. The widely studied HJM-model \cite{hjm} has been further developed by \cite{m} (and references cited therein) to account for stochastic volatility.  Amin and Ng \cite{an} studied the market data of Eurodollar options to obtain the implied forward rate volatility and Bouchaud et. al. \cite{lf} analyzed the future contracts for the forward rates. Both references concluded that many features of the market, and in particular the (stochastic) volatility of forward rate curve, could not be fully explained in the HJM-framework. 

The model for the forward rates proposed in \cite{beb1} is a field theoretic generalization of the HJM-model, and so it is natural to extend the field theory model to account for stochastic volatility of the forward rates.  In contrast to quantum field theory, the formulation of the forward rates as a stochastic string \cite{s1},\cite{s2} cannot be extended to the case where volatility is stochastic due to nonlinearites inherent in the problem.

The forward rates are the collection of interest rates for a contract entered into at time $t$ for an overnight loan at time $x>t$. At any instant $t$, there exists in the market forward rates for a duration of $T_{FR}$ in the future; for example, if $t$ refers to present time $t_0$, then one has forward rates from $t_0$ till time $t_0+T_{FR}$ in the future. In the market, $T_{FR}$ is about $30$ years, and hence we have $T_{FR}> 30$
years. In general, at any time $t$, all the forward rates exist till time
$t+T_{FR}$ \cite{beb1}. The forward rates at time $t$ are denoted by
$f(t,x)$, with $t<x<t+T_{FR}$, and constitute the {\bf forward rate curve}.

Since at any instant $t$ there are infinitely many forward rates, it resembles a (non-relativistic) quantum string. Hence we need an infinite number of independent variables to describe its random evolution.  The generic quantity describing such a system is a quantum 
field \cite{zj}. For modeling the forward rates and Treasury Bonds, we consequently need to study a two-dimensional quantum field on a finite Euclidean domain.

We consider the forward rates $f(t,x)$ to be a {\bf quantum field}; that
is, $f(t,x)$ is taken to be an {\bf independent} random variable for {\bf each} $x$ and {\bf each} $t$. For notational simplicity we consider both $t$ and  $x$
continuous, and discretize these parameters only when we need to discuss the time evolution of the system is some detail.

In Section 2, we briefly review the quantum field theoretic formulation given in \cite{beb1} of the forward rates with deterministic volatilty. In Section 3 the case of stochastic volatility is anlayzed, and which can be done in {\bf two different ways}. {\bf Firstly}, volatility can be considered to be a function of the (stochastic) forward rates, and {\bf secondly} volatility can be considered to be an independent quantum field. Both these cases are analyzed. The resulting theories are seen to be highly non-trivial non-linear quantum field theories. 

In Section 4 the underlying state space and operators of the forward rates quantum field is defined. In particular the generator of infinitesimal time evolution of the forward rates, namely the Hamiltonian, is derived for the two cases of stochastic volatility. In Section 5 the Hamiltonian for the forward rates with stochastic volatility is derived. In Section 6 a Hamiltonian formulation of the condition of no arbitrage is derived. In Section 7  the no arbitrage constraint for the case of stochastic volatility is solved exactly using the Hamiltonian formulation. And lastly, in Section 8 the results obtained are discussed, and some remaining issues are addressed.

\section {Lagrangian for Forward Rates with Deterministic Volatility}

We first briefly recapitulate the salient features of the field theory of the forward rates with deterministic volatility \cite{beb1}.

\begin{figure}[h]
  \begin{center}
    \epsfig{file=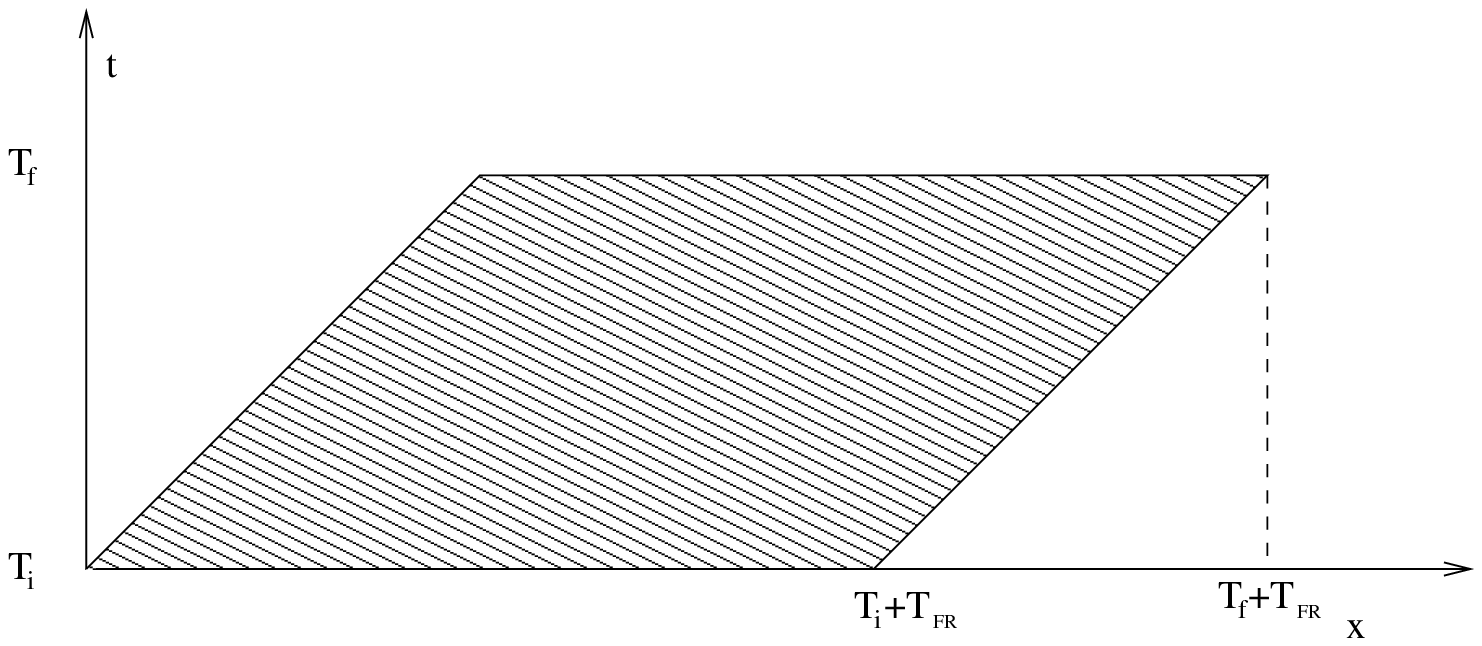,width=11cm}
    \caption{Domain ${\cal P}$ of the Forward Rates}
    \label{frc}
  \end{center}
\end{figure}

For the sake of concreteness, consider the forward rates starting from some initial time $T_i$
to a future time $t=T_f$. Since all the forward rates $f(t,x)$ are always for the future,
we have $x>t$; hence the quantum field $f(t,x)$ is defined on the domain in the
shape of a parallelogram ${\cal P}$ that is bounded by parallel lines
$x=t$ and $x=T_{FR}+t$ in the maturity direction , and by the lines $t=T_i$ and $t=T_f$ in the time direction, as shown in Figure (\ref{frc}). Every point inside the domain ${\cal P}$ represents an independent integration variable $f(t,x)$.

The field theory interpretation of the evolution of the forward rates, as expressed in the domain ${\cal P}$, is that of a (non-relativistic) quantum string moving with unit velocity in the $x$ (maturity) direction.

Since we know from the HJM-model that the forward rates have a drift velocity
$\alpha(t,x)$ and volatitility $\sigma(t,x)$, these have to appear
directly in the Lagrangian for the forward rates. To define a Lagrangian, we firstly need a kinetic term, denoted by ${\cal L}_{\mathrm{kinetic}}$, that is necessary to have a standard time evolution for the forward rates. 

We need to introduce another term to constrain the change of shape of the forward rates in the maturity direction. The analogy of this in the case of an ordinary string is a potential term in the Lagrangian which attenuates sharp changes in the shape of the string, since the shape of the string stores potential energy. To model a similar property for the forward rates  we cannot use a simple tension-like term $\displaystyle (\frac{\partial f}{\partial x})^2$  in the Lagrangian since, as we will show in Section 7, this term is ruled out by the condition of no arbitrage. 

The no arbitrage condition requires that the forward rates Lagrangian contain higher order derivative terms, essentially a term of the form $\displaystyle (\frac{\partial^2 f}{\partial x \partial t})^2$; such string systems have been studied in \cite{p} and are said to be strings with finite rigidity. Such a term yields a term in the forward rates Lagrangian, namely ${\cal L}_{\mathrm{rigidity}}$,  with a new parameter $\mu$; the rigidigy of the forward rates is then given by $\displaystyle \frac{1}{\mu^2}$ and quantifies the strength of the
fluctuations of the forward rates in the time-to-maturity
direction $x$. In the limit of
$\mu \rightarrow 0$, we recover (upto some rescalings) the HJM-model, and which corresponds to an infinitely rigid string.  The action for the forward rates is  given by
\begin{eqnarray}
\label{sf}
S[f]&=&\int_{T_i}^{T_f}dt\int_{t}^{t+T_{FR}}dx {\cal L}[f]\\
&\equiv& \int_{\cal P}{\cal L}[f]
\end{eqnarray}
with the Lagrangian density ${\cal L}[f]$ given by
\begin{eqnarray}
\label{lf}
{\cal L}[f]&=&{\cal L}_{\mathrm{kinetic}}[f]+{\cal
L}_{\mathrm{rigidity}}[f]\\
&=&-\frac{1}{2}\Big [\Big \{\frac{\frac{\partial
f(t,x)}{\partial t}- \alpha(t,x)}{\sigma(t,x)}\Big\}^2
+ \frac{1}{\mu^2}\Big \{ \frac{\partial}{\partial x}\Big (\frac{\frac{\partial
f(t,x)}{\partial t}- \alpha(t,x)}{\sigma(t,x)}\Big)\Big\}^2\Big]\nonumber\\
\label{rngf}
&&-\infty \leq f(t,x) \leq +\infty
\end{eqnarray}

The presence of the second term in the action given in eq.(\ref{lf}) is not ruled out by no arbitrage \cite{lf}, and an empirical study \cite{b2} provides strong evidence for this term 
in the evolution of the forward rates. 

In summary, we see that the forward rates behave like a quantum string, with a time and space dependent drift velocity $\alpha(t,x)$, an effective mass given by $\displaystyle
\frac{1}{\sigma(t,x)}$, and string rigidity proportional to $\displaystyle
\frac{1}{\mu^2}$.

Since the field theory is defined on a finite domain ${\cal P}$ as shown in Figure. \ref{frc}, we need to specify the boundary conditions on all the four boundaries of the finite parallelogram ${\cal P}$. 
\begin{itemize}
\item
{\bf Fixed (Dirichlet) Initial and Final Conditions}\\
The initial and final (Dirichlet) conditions in the time direction are given by
\begin{eqnarray}
\label{dif}
 T_i(T_f)<&x&<T_i(T_f)+T_{FR}~:~f(T_i,x),~f(T_f,x)\\
					&:&\mathrm{specified~initial~ and~final~forward ~rate~curves}\nonumber
\end{eqnarray}
\item
{\bf Free (Neumann) Bounday Conditions}\\
To specify the boundary condition in the maturity direction, one needs to analyze the action given in eq.(\ref{sf}) and impose the condition that there be no surface terms in the action. A straightforward analysis yields the following version of the Neumann condition
\begin{eqnarray}
\label{fbx}
T_i<t<T_f,&& \frac{\partial}{\partial x}\Big (\frac{\frac{\partial
f(t,x)}{\partial t}- \alpha(t,x)}{\sigma(t,x)}\Big)=0\\
						&:& x=t~\mathrm{or}~x=t+T_{FR}
\end{eqnarray}
\end{itemize}
The quantum field theory of the forward rates is defined by the Feynman path integral by integrating over all configurations, and yields
of $f(t,x)$ and 
\begin{eqnarray}
\label{zff}
Z&=&\int Df e^{S[f]}\\
\int Df&\equiv&\prod_{(t,x)\epsilon {\cal P}}\int_{-\infty}^{+\infty}df(t,x)
\end{eqnarray}

Note that $e^{S[f]}$/ Z is the
probability for different field configurations to occur when the functional
integral over $f(t,x)$ is performed.

\section {Lagrangian for Forward Rates with Stochastic Volatility}

To render the volatility function $\sigma(t,x)$ stochastic, in the formalism of quantum field theory, requires that we elevate $\sigma(t,x)$ from a deterministic function into random function, namely into a quantum field. There are essentially two ways of elevating volatility to a stochastic quantity, namely to either (a) consider it be a function of the forward rate $f(t,x)$, or else (b) to consider it to be another independent quantum field $\sigma(t,x)$. We study both these possibilities.

\subsection{Volatility a Function of the Forward Rates}

We consider the first case where volatility is rendered stochastic by making it a function of the forward rates \cite{an}. The standard models using this approach consider that volatility is given by
\begin{eqnarray}
\label{fvol}
\sigma(t,x, f(t,x))&=&\sigma_0(t,x) f^\nu(t,x)\\
\sigma_0(t,x)	&&:\mathrm{Deterministic~function}
\end{eqnarray}

Since volatility $\sigma(t,x) >0$, we must have $f(t,x) >0$. Hence, in contrast to eq.(\ref{rngf}), we have
\begin{eqnarray}
f(t,x)=f_0 e^{\phi(t,x)}>0~;~-\infty \leq \phi(t,x) \leq +\infty
\end{eqnarray}
Having $f(t,x)>0$ is a major advantage of the model since in the financial markets forward rates are always positive.  In the limit of $\mu\rightarrow 0$, the following HJM-models are covered by eq.(\ref{fvol}), and these models have been discussed from an empirical point of view in \cite{an}.

\begin{enumerate}
\item Ho and Lee (1986) Model : $\sigma(t,x, f(t,x))= \sigma_0$
\item CIR (1985)						: $\sigma(t,x, f(t,x))= \sigma_0 f^{\frac{1}{2}}(t,x)$
\item Courtadon(1982) :$\sigma(t,x, f(t,x))= \sigma_0 f(t,x)$	
\item Vasicek (1977) :$\sigma(t,x, f(t,x))= \sigma_0 \exp(-\lambda(x-t))$
\item Linear Proportional HJM (1992) : $\sigma(t,x, f(t,x))= [\sigma_0+\sigma_1(x-t)]f(t,x)]$
\end{enumerate}

How do we generalize the Lagrangian given in eq.(\ref{lf}) to case where the forward rates are always positive? We interpret the Lagrangian given in eq.(\ref{lf}) to be an approximate one that valid only if all the forward rates are close to some fixed value $f_0$. We then have
\begin{eqnarray}
\frac{\partial f(t,x)}{\partial t }&=&  f_0 e^{\phi(t,x)}\frac{\partial \phi(t,x)}{\partial t}\\
&\simeq& f_0 \frac{\partial\phi(t,x)}{\partial t}+O(\phi^2)
\end{eqnarray}
Hence we make the following mapping
\begin{eqnarray}
\frac{\partial f(t,x)}{\partial t }\rightarrow f_0\frac{\partial \phi(t,x)}{\partial t}
\end{eqnarray}
Eq.(\ref{lf}) then generalizes to
\begin{eqnarray}
\label{lagphi}
{\cal L}[\phi]&=&{\cal L}_{\mathrm{kinetic}}[\phi]+{\cal
L}_{\mathrm{rigidity}}[\phi]\\
&=&-\frac{1}{2}\Big [\Big \{\frac{f_0\frac{\partial
\phi(t,x)}{\partial t}- \alpha(t,x)}{\sigma_0(t,x) e^{\nu\phi(t,x)}}\Big\}^2
+ \frac{1}{\mu^2}\Big \{ \frac{\partial}{\partial x}\Big (\frac{f_0\frac{\partial
\phi(t,x)}{\partial t}- \alpha(t,x)}{\sigma_0(t,x) e^{\nu\phi(t,x)}}\Big)\Big\}^2\Big]\nonumber
\end{eqnarray}
We will show later -- in deriving the Hamiltonian -- that the system needs a non-trivial integration measure. We hence define the theory by the Feynman path integral
\begin{eqnarray}
\label{zff}
Z&=&\int D\phi ~f^{-\nu}e^{S[\phi]}\\
\int D\phi~f^{-\nu}&\equiv&\prod_{(t,x)\epsilon {\cal P}}\int_{-\infty}^{+\infty} d\phi(t,x)~f^{-\nu}(t,x)
\end{eqnarray}
The boundary conditions given for $f(t,x)$ in eqs.(\ref{dif}) and (\ref{fbx}) continue to hold for stochastic volatility Lagrangian given in eq.(\ref{lagphi}) .
 
\subsection{Volatility as an Independent Quantum Field}

We consider the second case where volatility $\sigma(t,x)$ is taken to be an {\bf independent} (stochastic) quantum field. Since one can only measure the effects of volatility on the forward rates, all the effects of stochastic volatility will be manifested only via the behaviour of the forward rates. 

For simplicity, we consider the forward rate to be a quantum field as given in eq.(\ref{rngf}) with 
\begin{eqnarray}
f(t,x):~~~-\infty \leq f(t,x) \leq +\infty
\end{eqnarray}
 
Since the volatility function $\sigma(t,x)$ is always positive, that is, $\sigma(t,x)>0$, we introduce an another quantum field $h(t,x)$ by the following relation (the minus sign is taken for notational convinience).
\begin{eqnarray}
\sigma(t,x)=\sigma_0 e^{-h(t,x)},~~-\infty \leq h(t,x) \leq +\infty
\end{eqnarray}

The system now consists of {\bf two interacting quantum fields}, namely $f(t,x)$ and $h(t,x)$. The interacting system's Lagrangian should have the following features.
\begin{itemize}
\item A parameter $\xi$ that quantifies the extent to which the field $h(t,x)$ is not deterministic. A limit of $\xi \rightarrow 0$ would, in effect, `freeze' all the fluctuations of the field $h(t,x)$, and reduce it to a deterministic function.
\item A parameter $\kappa$  to control the fluctuations of $h(t,x)$ in the maturity direction similar to the parameter $\mu$ that controls the fluctuations of the forward rates $f(t,x)$ in the maturity direction $x$.
\item A parameter $\rho$ with $-1 \leq \rho \leq +1$ that quantifies the correlation of the forward rates' quantum field $f(t,x)$ with the volatility quantum field 
$h(t,x)$.
\item A drift term for volatility, namely $\beta(t,x)$ -- which is analogous to the drift term 
$\alpha(t,x)$ for the forward rates. 
\end{itemize}

The Lagrangian for the interacting system is not unique; there is a wide variety of choices that one can make to fullfil all the conditions given above. A possible Lagrangian for the interacting system, written by analogy with the Lagrangian for the case of stochastic volatility for a single security \cite{beb2}, is given by
\begin{eqnarray}
\label{stochasI}
{\cal L} &=& -\frac{1}{2(1-\rho^2)} \left(\frac{\frac{\partial f}{\partial t} - \alpha}{\sigma} - \rho \frac{\frac{\partial h}{\partial t} - \beta}{\xi}\right)^2 -
\frac{1}{2} \left(\frac{\frac{\partial h}{\partial t} -
\beta}{\xi}\right)^2 \nonumber\\
&& - \frac{1}{2\mu^2}\left(\frac{\partial}{\partial x}
\left(\frac{\frac{\partial f}{\partial t} - \alpha}{\sigma} \right)\right)^2
-\frac{1}{2 \kappa^2}
\left(\frac{\partial}{\partial x}\left(\frac{\frac{\partial
h}{\partial t} - \beta}{\xi}\right)\right)^2  
\end{eqnarray}
with action
\begin{eqnarray}
\label{sfh}
S[f,h]=\int_{{\cal P}}~{\cal L}
\end{eqnarray}

We need to specify the boundary conditions for the interacting system. The initial and final conditions for the forward rates $f(t,x)$ given in eq.(\ref{dif}) continue to hold for the interacting case, and are similarly given for the volatility field as the following.
\begin{itemize}
\item
{\bf Fixed (Dirichlet)Initial and Final Conditions}\\
The initial value is specified from data, that is
\begin{eqnarray}
T_i(T_f)<&x&<T_i(T_f)+T_{FR},~~\sigma(T_i,x),\sigma(T_f,x) \\
		&:&\mathrm{specified~initial~and ~final~volatility~curves}\nonumber
\end{eqnarray}
\end{itemize}
The boundary condition in the $x-$direction for the forward rates $f(t,x)$ -- as given in eq.(\ref{fbx}) -- continues to hold for the interacting case, and for volatility field is similarly given by
\begin{itemize}
\item
{\bf Free (Neumann) Boundary Conditions}
\begin{eqnarray}
\label{hx}
T_i<t<T_f &;&~\frac{\partial}{\partial x}\Big (\frac{\partial
h(t,x)}{\partial t}- \beta(t,x)\Big)= 0\\
						: x=t~\mathrm{or}~x&=&t+T_{FR}
\end{eqnarray}
\end{itemize}
On quantizing the volatility field $\sigma(t,x)$ the boundary condition for the forward rate $f(t,x)$  given in eq.(\ref{fbx}) is rather unusual. On solving the no arbitrage condition, we will find that $\alpha$ is a (quadratic) functional of the volatility field $\sigma(t,x)$; hence the boundary condition eq.(\ref{fbx}) is a form of {\bf interaction} between the $f(t,x)$ and $\sigma(t,x)$ fields. 

We need to define the  integration  measure for the quantum field $h(t,x)$; the derivation of the Hamiltonian for the system dictates the following choice for the measure, namely   
\begin{eqnarray}
\int DfD\sigma^{-1} &=&\prod_{(t,x)\epsilon {\cal P}}\int_{-\infty}^{+\infty}df(t,x)d\sigma^{-1}(t,x)\\
\label{pim}
						&=&\prod_{(t,x)\epsilon {\cal P}}
						\int_{- \infty}^{+\infty}df(t,x)dh(t,x)e^{h(t,x)}
\end{eqnarray}

The partition function of the quantum field theory for the forward rates with stochastic volatility is defined by Feynman path integral as
\begin{eqnarray}
\label{zfh}
Z=\int DfD\sigma^{-1} e^{S[f,h]}
\end{eqnarray}

The (observed) market value of a financial instrument,  say ${\cal O}[f,h]$,  is expressed as the {\bf average value} of the instrument -- denoted by $<{\cal O}[f,h]>$ --  taken over all possible values  of the quantum fields $f(t,x)$ and $h(t,x)$, with the probability density given by the (appropriately normalized) action. In symbols
\begin{eqnarray}
\label{exval}
<{\cal O}[f,h]>=\frac{1}{Z}\int DfD\sigma^{-1} ~{\cal O}[f,h] e^{S[f,h]}
\end{eqnarray}

We consider the limit of the volatility being reduced to a deterministic function. For this limit we have $\xi, \rho$ and $\kappa \rightarrow 0$. The kinetic term of the $h(t,x)$ field in the action given in eq.(\ref{sfh}) has the limit (upto irrelevant constants) 
\begin{eqnarray}
\lim_{\xi\rightarrow 0}\prod_{t,x \epsilon {\cal P}} \exp\Big \{ -\frac{1}{2}\int_{{\cal P}}  \left(\frac{\frac{\partial h}{\partial t} -
\beta}{\xi}\right)^2\Big \} \rightarrow \prod_{t,x \epsilon {\cal P}}\delta(\frac{\partial h}{\partial t} -\beta)
\end{eqnarray}
which implies that
\begin{eqnarray}
 <\sigma(t,x)>&=&\sigma_0<e^{-h(t,x)}>\\
			&=&\sigma_0 \exp\big \{-\int_{t_0}^t~dt' \beta(t',x)\big \}+ O(\xi,\kappa,\rho) 
\end{eqnarray}

\sect{Hamiltonian and State Space}

The Feynman path integral formulation given in eqs. (\ref{zff}) and (\ref{zfh}) is useful for calculating the expectation values of quantum fields. To study questions related to the time evolution of quantities of interest, one needs to derive the Hamiltonian for the system from its Lagrangian. Note the route that we are following is opposite to the one taken in \cite{beb2} where the Lagrangian for a stock price with stochastic volatility was derived starting from its Hamiltonian.

The state space of a field theory is a linear vector space -- denoted by ${\cal V}$ -- that consists of functionals of the field configurations at some fixed time $t$.  (A brief discussion of the state space is given in \cite{beb2}.) The dual space of ${\cal V}$ -- denoted by ${\cal V}_{\mathrm{Dual}}$ -- consists of all linear mappings from ${\cal V}$ to the complex numbers, and is also a linear vector space. Let an element of ${\cal V}$ be denoted by $|g>$ and an element of ${\cal V}_{\mathrm{Dual}}$ by $<p|$; then $<p|g>$ is a complex number. We will refer to both ${\cal V}$ and ${\cal V}_{\mathrm{Dual}}$ as the state space of the system. The Hamiltonian ${\cal H}$  is an  operator -- the quantum analog of energy -- that is an element of the tensor product space ${\cal V} \otimes {\cal V}_{\mathrm{Dual}}$. The matrix elements of ${\cal H}$ are complex numbers, and given by $<p|{\cal H}|g>$.

In this section, we study the  features of the state space and Hamiltonian for the forward rates. For notational brevity, we consider the forward rates  quantum field $f(t,x)$ to stand for both the quantum fields $f(t,x)$ and $h(t,x)$. Since the Lagrangian for the forward rates given in eq.(\ref{stochasI}) has only first order derivatives in time, an infinitesimal generator, namely the Hamiltonian ${\cal H}$ exists for it. Obtaining the Hamiltonian for the forward rates is a complicated exercise due to the non-trivial structure of the underlying domain ${\cal P}$. In particular, the forward rates quantum field will be seen to have a distinct state space ${\cal V}_t$ for every instant $t$.

\begin{figure}[h]
\begin{center}
\epsfig{file=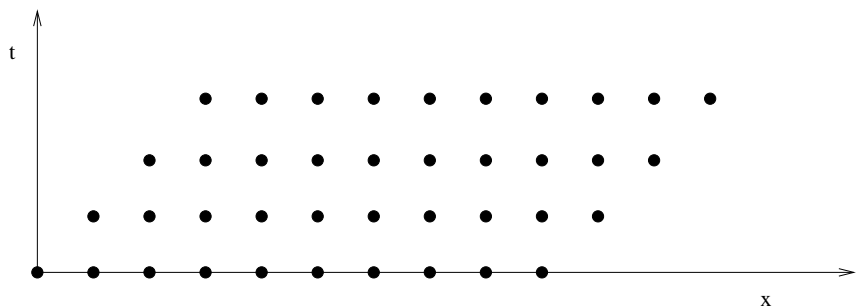,height=4cm}
\caption{Lattice in Time and Maturity Directions}
\label{txlat}
\end{center}
\end{figure}

For greater clarity, we discretize both time and maurity time into a finite lattice, with lattice spacing in both directions taken to be $\epsilon$. (For a string moving with velocity $v$, the maturity lattice would have spacing of $v\epsilon$.) On the lattice, the minimum time for futures contract is time $\epsilon$; for  most applications $\epsilon=$one day. The points comprising the discrete domain $\tilde{\cal P}$ are shown in Figure \ref{txlat}. 

The  discrete domain  $\tilde{\cal P}$ is given by
\begin{eqnarray}
\label{disf}
(t,x)&\rightarrow& \epsilon(n,l)~;~n,l:~\mathrm{integers}\\
(T_i,T_f,T_{FR}) &\rightarrow& \epsilon(N_i,N_f,N_{FR})\\
\mathrm{Lattice}~\tilde{\cal P}&=&\{(n,l)|N_i\leq n \leq N_f~;~n\leq l \leq {(n+N_{FR})}\}\\
f(t,x)&\rightarrow& f_{n,l}\\
\label{dtd}
\frac{\partial f(t,x)}{\partial t}&\simeq & \frac{f_{n+1,l}-f_{n,l}}{\epsilon};~\frac{\partial f(t,x)}{\partial x} \simeq \frac{f_{n,l+1}-f_{n,l}}{\epsilon}
\end{eqnarray}

The partition function is now given by a finite multiple integral, namely
\begin{eqnarray}
\label{disz}
Z&=&\prod_{(n,l) \epsilon \tilde{\cal P}} \int df_{n,l}e^{S[f]}\\
S&=&\sum_n S(n)
\end{eqnarray}

\begin{figure}[h]
\epsfig{file=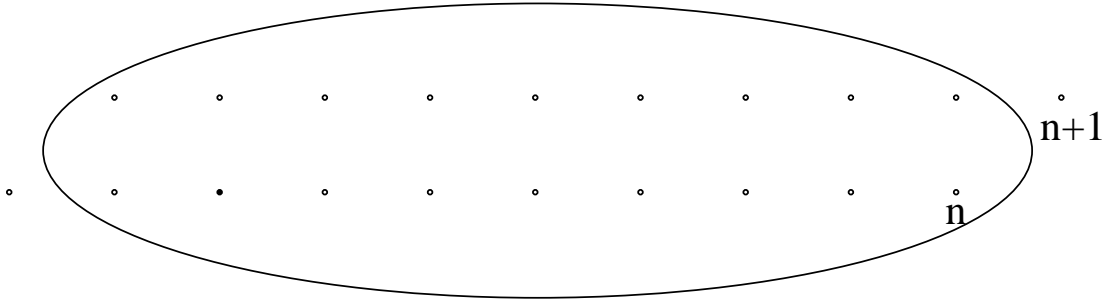,height=4cm}
\caption{Two Consecutive Time Slices for $t=n\epsilon$ and $t=(n+1)\epsilon$}
\label{2lat}
\end{figure}

Consider two adjacent time slices labelled by $n$ and $n+1$, as shown in Figure \ref{2lat}. $S(n)$ is the action connecting the forward rates of these two time slices. 

As can be seen from Figure \ref{2lat}, for the two time slices there is a mismatch of the $2$-lattice sites on the edges, namely, lattice sites $(n,n)$ at time $n$ and $(n+1,n+1+N_{FR})$ at time $n+1$ are not in common. We isloate the un-matched variables and have the following

\begin{eqnarray}
\label{coords}
\mathrm{Variables~at~time~n}:&& \nonumber\\
\{f_{n,n},\tilde{F}_n \}~; && \tilde{F}_n\equiv \{f_{n,l}|n \leq l \leq n+N_{FR}\}\\
\mathrm{Variables~at~time~(n+1)}:&& \nonumber \\
\{F_n,f_{n+1,n+1+N_{FR}} \}~;&& F_n \equiv \{f_{n+1,l}|n+1 \leq l \leq n+1+N_{FR}\}
\end{eqnarray}
Note that although the variables $F_n$ refer to time $n+1$, we label it with earlier time $n$ for later convinience. From Figure \ref{2lat} we see that both {\bf sets of variables} $F_n$ and $\tilde{F}_n$ cover the {\bf same} lattice sites in the maturity direction, namely 
$n+1 \leq l \leq n+N_{FR}$, and hence have the same number of forward rates, namely $N_{FR}-1$. The Hamiltonian will be expressable solely in terms of these variables. 

From the discretized time derivatives defined in eq.(\ref{dtd})  the discretized action $S(n)$ contains terms that couple only the common points in the lattice for the two time slices, namely the variables belonging to the sets $\tilde{F}_n;F_n$. 
We hence have for the action
\begin{eqnarray}
\label{disac}
S(n)&=&\epsilon \sum_{\{l\}} {\cal L}_{n}[f_{n,l},f_{n+1,l}]\\
\label{dscac}
&=&\epsilon \sum_{\{l\}} {\cal L}_{n}[\tilde{F}_n;F_n]
\end{eqnarray}

\begin{figure}[h]
\begin{center}
\epsfig{file=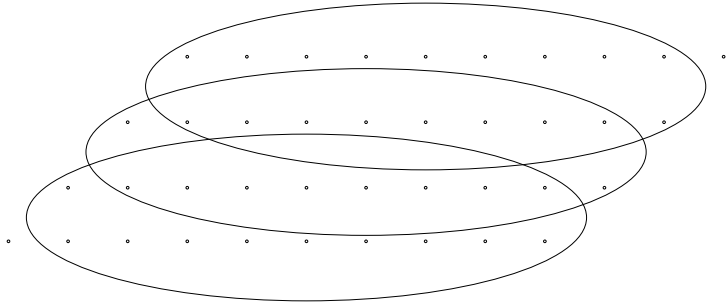,height=4cm}
\caption{Reconstructing the Lattice from the Two Time Slices}
\label{nlat}
\end{center}
\end{figure}

As shown is in Figure \ref{nlat}, the action for the entire domain $\tilde{\cal P}$ shown in Figure \ref{txlat} can be  constructed by repeating the construction given in Figure \ref{2lat} and summing over the action $S(n)$ over all time $N_i \leq n \leq N_f$.

The Hamiltonian of the forward rates is an operator that acts on the state space of states of  the forward rates; we hence need to determine the co-ordinates of its state space. 

Consider again the two consecutive time slices $n$ and $n+1$ given in Figure \ref{2lat}. We interpret the forward rates for two adjacent instants, namely $\{f_{n,n},\tilde{F}_n \}$ and $\{F_n,f_{n+1,n+1+N_{FR}} \}$ given in  eq .(\ref{coords}) -- and which appear in the action eq.(\ref{disac}) --  as the co-ordinates of the state spaces ${\cal V}_n$ and ${\cal V}_{n+1}$ respectively. 

For every instant of time $n$ there is a distinct state space ${\cal V}_n$, and its dual ${\cal V}_{\mathrm{Dual},n}$. The co-ordinates of the  state spaces  ${\cal V}_n$ and ${\cal V}_{n+1}$
 are given by the tensor product of the space of state for every maturity point $l$, namely
\begin{eqnarray}
<\tilde{f}_n|&=& \bigotimes_{n \leq l \leq n+N_{FR}} <f_{n,l}|\equiv <f_{n,n}|<\tilde{F}_n|\\
				&:&~\mathrm{co-ordinate~ state ~of~ {\cal V}_{\mathrm{Dual},n}} \nonumber\\
\label{stvec}
|f_{n+1}>&=&\bigotimes_{(n+1) \leq l \leq n+1+N_{FR}}|f_{n+1,l}> \equiv |F_n>|f_{n+1,n+1+N_{FR}}>\\
				&:&~\mathrm{co-ordinate~ state~ of~ {\cal V}_{n+1}} \nonumber
\end{eqnarray}
The state vector $|F_n>$ belongs to the space space ${\cal V}_{n+1}$, but we reinterpret it as corresponding to  the state space ${\cal F}_n$ at earlier time $n$. This interpretation allows us to study the system instantaneously using the Hamiltonian formalism.

The state space ${\cal V}_n$ consists of all possible functions of $N_{FR}$ forward rates 
$\{f_{n,n},\tilde{F}_n\}$. The state spaces ${\cal V}_n$ differ for different $n$ by the fact that a different set of forward rates comprise its set of independent variables. 

Although the state spaces ${\cal V}_n$ and ${\cal V}_{n+1}$ are not the identical, there is an intersection of these two spaces, namely ${\cal V}_n \cap {\cal V}_{n+1}$ that covers the same interval in the maturity direction, and is coupled by the action $S(n)$. The intersecion yields a state space, namely ${\cal F}_n$, on which the Hamiltonian evolution of the forward rates takes place. In symbols, we have
\begin{eqnarray}
{\cal V}_{n+1}&=&{\cal F}_n\otimes|f_{n+1,n+1+N_{FR}}>\\
{\cal V}_{\mathrm{Dual},n}&=& <f_{n,n}| \otimes {\cal F}_{\mathrm{Dual},n}\\
{\cal H}_n &:& ~{\cal F}_n \rightarrow {\cal F}_{n}~~\Rightarrow~{\cal H}_n \in ~{\cal V}_{\mathrm{Dual},n} \otimes  {\cal V}_{n+1}
\end{eqnarray}
The Hamiltonian ${\cal H}_n$ is an element on the tensor product space spanned by the operators 
$|F_n><\tilde{F}_n|$, namely the space of operators given by 
${\cal F}_n\otimes {\cal F}_{\mathrm{Dual},n}$.

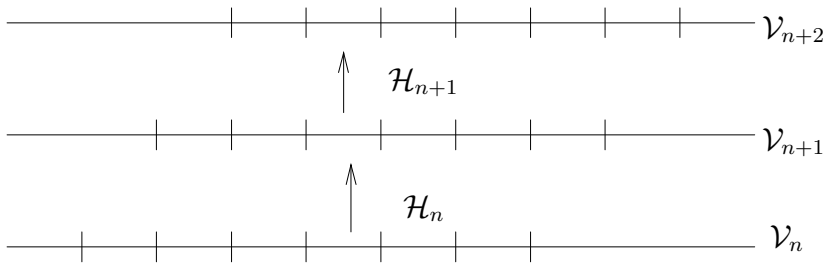
\begin{figure}[h]
\begin{center}
\setlength{\unitlength}{0.00087489in}
\begingroup\makeatletter\ifx\SetFigFont\undefined%
\gdef\SetFigFont#1#2#3#4#5{%
  \reset@font\fontsize{#1}{#2pt}%
  \fontfamily{#3}\fontseries{#4}\fontshape{#5}%
  \selectfont}%
\fi\endgroup%
{\renewcommand{\dashlinestretch}{30}
\begin{picture}(4602,1572)(0,-10)
\path(462,192)(462,12)
\path(912,192)(912,12)
\path(1362,192)(1362,12)
\path(1812,192)(1812,12)
\path(2262,192)(2262,12)
\path(2712,192)(2712,12)
\path(3162,192)(3162,12)
\path(912,867)(912,687)
\path(1362,867)(1362,687)
\path(1812,867)(1812,687)
\path(2262,867)(2262,687)
\path(2712,867)(2712,687)
\path(3162,867)(3162,687)
\path(3612,867)(3612,687)
\path(1362,1542)(1362,1362)
\path(1812,1542)(1812,1362)
\path(2262,1542)(2262,1362)
\path(2712,1542)(2712,1362)
\path(3162,1542)(3162,1362)
\path(3612,1542)(3612,1362)
\path(4062,1542)(4062,1362)
\path(12,102)(4512,102)
\path(12,777)(4512,777)
\path(12,1452)(4512,1452)
\path(2112.000,477.000)(2082.000,597.000)(2052.000,477.000)
\path(2082,597)(2082,192)
\path(2067.000,1152.000)(2037.000,1272.000)(2007.000,1152.000)
\path(2037,1272)(2037,912)
\put(2397,282){\makebox(0,0)[lb]{\smash{{{\SetFigFont{12}{14.4}{\rmdefault}{\mddefault}{\updefault}${\cal H}_n$}}}}}
\put(2307,1047){\makebox(0,0)[lb]{\smash{{{\SetFigFont{12}{14.4}{\familydefault}{\mddefault}{\updefault}${\cal H}_{n+1}$}}}}}
\put(4602,102){\makebox(0,0)[lb]{\smash{{{\SetFigFont{12}{14.4}{\rmdefault}{\mddefault}{\updefault}${\cal V}_n$}}}}}
\put(4557,687){\makebox(0,0)[lb]{\smash{{{\SetFigFont{12}{14.4}{\rmdefault}{\mddefault}{\updefault}${\cal V}_{n+1}$}}}}}
\put(4557,1362){\makebox(0,0)[lb]{\smash{{{\SetFigFont{12}{14.4}{\rmdefault}{\mddefault}{\updefault}${\cal V}_{n+2}$}}}}}
\end{picture}
}
\caption{Hamiltonians ${\cal H}_n $ propagating the space of Forward Rates ${\cal V}_n$ }
\end{center}
\label{vh}
\end{figure}

The vector spaces ${\cal V}_n$ and the Hamiltonian ${\cal H}_n$ acting on these spaces is shown in Figure \ref{vh}. 

Note that both the states $|F_n>$ and $<\tilde{F}_n|$ belong to the same state space 
${\cal F}_n$, and we use twiddle to indicate that the two states are {\bf different}; in contrast, for example, the two states $|f>$ and $<f|$ indicate that one state is the dual 
of the other.

As one scans through all  possible values for the forward rates $f$ and $\tilde{f}$, one obtains a complete basis for the state space ${\cal V}_n$. In particular, the resolution of the identity operator for ${\cal V}_n$ -- denoted by ${\cal I}_n$ -- is a reflection that the basis states are complete, and is given by \cite{beb2}
\begin{eqnarray}
\label{comeq}
{\cal I}_n&=&\prod_{n \leq l \leq n+N_{FR}} \int df_{n,l}|f_n><f_n|\\
           &\equiv & \int df_{n,n}~ d \tilde{F}_n ~|f_{n,n};\tilde{F}_n><f_{n,n};\tilde{F}_n|
\end{eqnarray}

The Hamiltonian of the system ${\cal H}$ is defined by the Feynman formula (upto a normalization), from eq.(\ref{disac}), as
\begin{eqnarray}
\label{ffor}
\rho_n e^{\epsilon \sum_{\{l\}} {\cal L}_{n}[f_{n,l},f_{n+1,l}]}&=&<f_{n,n},\tilde{F}_n|e^{-\epsilon {\cal H}_n}|F_n,f_{n+1,n+1+N_{FR}}>
\end{eqnarray}
where in general $\rho_n$ is a field-dependent measure term.
Using the property of the discrete action given in eq.(\ref{dscac}), we  have 
\begin{eqnarray}
\label{ffor}
\rho_n e^{\epsilon \sum_{\{l\}} {\cal L}_{n}[F_n,\tilde{F}_n]}&=&<f_{n,n},\tilde{F}_n|e^{-\epsilon {\cal H}_n}|F_n,f_{n+1,n+1+N_{FR}}>\\
\label{hamilt}
							&=&<\tilde{F}_n|e^{-\epsilon {\cal H}_n}|F_n>
\end{eqnarray}

Equation (\ref{hamilt}) is the main result of this Section.

In going from eq.(\ref{ffor}) to eq.(\ref{hamilt}) we have used the fact that the action connecting time slices $n$ and $n+1$ does not contain the variables $f_{n,n}$ and 
$f_{n+1,n+1+N_{FR}}$ respectively. This leads to the result that the Hamiltonian ${\cal H}_n$ consequently  does not depend on these variables. 

The interpretation of eq.(\ref{hamilt}) is that the Hamiltonian ${\cal H}_n$ propagates the initial state $<\tilde{F}_n|$ in time $\epsilon$ to the final state $|F_n>$. 
Note the relation
\begin{eqnarray}
<f_{n,n},\tilde{F}_n|e^{-\epsilon {\cal H}_n}|F_n,f_{n+1,n+1+N_{FR}}>=
<\tilde{F}_n|e^{-\epsilon {\cal H}_n}|F_n>
\end{eqnarray}
shows that there is an asymmetry in the time direction, with the Hamiltonian being {\bf independent} of the {\it earliest} forward rate $f_{n,n}$ of the initial state and of the {\it latest} forward rate $f_{n+1,n+1+N_{FR}}$ of the final state. It is this {\bf asymmetry} in the propagation of the forward rates which yields the parallelogram domain ${\cal P}$ given in Figure \ref{txlat}, and reflects the asymmetry  that the forward rates $f(t,x)$ exist only for $x>t$.

For notational simplicity, we henceforth use continuum notation; in particular, the state space is labelled by ${\cal V}_t$, and state vector by $|f_t>$. The elements of the state space  of the forward rates ${\cal V}_t$ includes {\bf all} the financial instruments that are traded in the market at time $t$. In continuum notation, from eq.(\ref{stvec}), we have  that 
\begin{eqnarray}
|f_t>&=& \bigotimes_{t \leq x \leq t+T_{FR}}|f(t,x)> \\
\label{fcont}
|F_t>&=& \bigotimes_{t < x \leq t+T_{FR}}|f(t,x)> 
\end{eqnarray}
In continuum notation, the only difference between state vectors $|f_t>$ and $|F_t>$ is that, in eq.(\ref{fcont}), the point $x=t$ is excluded in the continuous tensor product.

The partition function $Z$ given in eq.(\ref{disz}) can be reconstructed from the Hamiltonian by recursively applying the procedure discussed for the two time slices. We then have, in continuum notation, that
\begin{eqnarray}
Z&=&\int Df e^{S[f]}\\
  &=& <f_{\mathrm{initial}}|{\cal T}\Big \{\exp(-\int_{T_i}^{T_f}{\cal H}(t)~dt)\Big \}
																	|f_{\mathrm{final}}>
\end {eqnarray}
where the symbol ${\cal T}$ in the equation above stands for time ordering the (non-commuting) operators in the argument, with the earliest time being placed to the left.

\subsection{Bond State Vectors}
The most important state vectors in finance are those of the coupon and zero coupon bonds.

Consider a risk-free zero coupon Treasury bond that matures at time $T$ with a payoff of $\$1$. The price of the bond at time $t<T$ is given by
\begin{eqnarray}
P(t,T)=e^{-\int_{t}^{T}f(t,x)dx}\equiv P[f;t,T]
\end {eqnarray}

The state vector $|P(t,T)>$ is an element of the state space ${\cal V}_t$. We write the bond state vector is as follows 
\begin{eqnarray}
P(t,T)&\equiv &<f_t|P(t,T)>\\
\label{zcb}
			&=& e^{-\int_{t}^{T}f(x)dx}
\end {eqnarray}

Another state vector is the coupon bond $|{\cal B}>$, with payoff of amount $c_l$ at time $T_l$,  with a final payoff of $L$ at time $T$. We then have that the state vector of the coupon bond is  linear superpostion of the zero coupon bonds and is given by
\begin{eqnarray}
|{\cal B}(t)>=\sum_l c_l|P(t,T_l)> + L|P(t,T)>
\end {eqnarray}

\section {Hamiltonian for the Forward Rates with Stochastic Volatility}
We have obtained a general expression for the Hamiltonian in terms of the action $S$ as given in eq.(\ref{hamilt}), and need to apply this formula to the case of the specific Lagrangian of the forward rates to obtain the explicit expression for its Hamiltonian ${\cal H}$. 

From eq.(\ref{hamilt}) we have the following
\begin{eqnarray}
\label{hamfh}
\rho_n e^{S(n)}&=& \rho_n  e^{\epsilon \sum_{\{l\}} {\cal L}_{n}[H_n,\tilde{H}_n;F_n,\tilde{F}_n]}\\
								&=&<\tilde{F}_n;\tilde{H}_n|e^{-\epsilon {\cal H}_n}|F_n;H_n> 
\end{eqnarray}
where we have explicitly included the volatility quantum field $h(t,x)$ in the equation above.

For notational simplicity, we consider the maturity direction $x$ to be continuous, and  consider only the time direction to be discrete. In the continuum notation, the subtleties of the variables at time $t$ and $t+\epsilon$ are accounted for by carefully analyzing the variables appearing on the {\bf boundaries} of the interval $[t \leq x\leq t+T_{FR}]$. We have, for the action $S(n)$ for time $t=n\epsilon$, the following 
\begin{eqnarray}
S(n) &=&\epsilon \int{\cal L}_n(t,x)\\
\int&\equiv&\int_{t}^{t+T_{FR}}~dx
\end{eqnarray}

\subsection{Hamiltonian for the Forward Rates with Stochastic Volatility}
As a warm-up exercise, we first obtain the Hamiltonian for the simpler case of the volatility being a function of the forward rates. Recall the Lagrangian for this system is given by
\begin{eqnarray}
\label{lphi}
{\cal L}[\phi]&=&{\cal L}_{\mathrm{kinetic}}[\phi]+{\cal
L}_{\mathrm{rigidity}}[\phi]\\
&=&-\frac{1}{2}\Big [\Big \{\frac{f_0\frac{\partial
\phi(t,x)}{\partial t}- \alpha(t,x)}{\sigma_0(t,x) e^{\nu\phi(t,x)}}\Big\}^2
+ \frac{1}{\mu^2}\Big \{ \frac{\partial}{\partial x}\Big (\frac{f_0\frac{\partial
\phi(t,x)}{\partial t}- \alpha(t,x)}{\sigma_0(t,x) e^{\nu\phi(t,x)}}\Big)\Big\}^2\Big]\nonumber
\end{eqnarray}
On discretizing the lagragian we obtain, using boundary condition eq.(\ref{fbx}), that
\begin{eqnarray}
\label{asf}
S(n)=\epsilon \int {\cal L}_n&=&-\frac{1}{2 \epsilon}\int A \Big (1-\frac{1}{\mu^2}\frac{\partial^2}{\partial x^2}\Big )A\\
 A&=&f_0\sigma_0^{-1}e^{-\nu\phi}(\phi_{t+\epsilon}-\phi_t-\epsilon f_0^{-1} \alpha)
\end{eqnarray}
We rewrite eq.(\ref{asf}) using Gaussian integration and obtain (ignoring henceforth irrelevant constants)
\begin{eqnarray}
\label{pl}
e^{S(n)}=\prod_{x}\int dp(x)\exp\big \{ -\frac{\epsilon}{2}\int p(x)D(x,x';t)p(x')+i\int p(x)A(x) \big \}\\
(1-\frac{1}{\mu^2}\frac{\partial^2}{\partial x^2})D(x,x';t)=\delta(x-x')~;~\mathrm{Neumann~boundary~conditions}\nonumber
\end{eqnarray}
An explicit derivation of the propagator $D(x,x';t)$ is given in the Appendix, and yields
\begin{eqnarray}
\label{prop}
D(x,x';t)=\frac{\mu}{2 \sinh(\mu T_{FR})} 
			 [\cosh(\mu T_{FR}&-&\mu |x-x'|) \nonumber\\
				+ \cosh(\mu T_{FR}&-&\mu (x+x'-2t))]
\end{eqnarray}

Let the measure term be defined as 
\begin{eqnarray}
\rho_n=\prod_{x}f^{-\nu}(n\epsilon,x)
\end{eqnarray}
Rescale $p(x)$ as follows (and which will effectively cancell the measure term)
\begin{eqnarray}
p\rightarrow f_0^{-1} \sigma_0 e^{\nu \phi}p
\end{eqnarray}
We then have
\begin{eqnarray}
\rho_n e^{S(n)}&=&\prod_{x}\int dp(x)\exp\Big \{
i\int p(x)(\phi_{t+\epsilon}-\phi_t-\epsilon f_0^{-1} \alpha)(x) \nonumber \\
&&-\frac{\epsilon}{2 f_0^{2} }\int \sigma_0 e^{\nu\phi}p(x) D(x,x';t)\sigma_0 e^{\nu \phi}p(x') \Big \}
\end{eqnarray}
Recall from eq.(\ref{hamilt}) that the Hamiltonian is defined by
\begin{eqnarray}  	
 \rho_n e^{S(n)}
  &=&<\phi_t|e^{-\epsilon {\cal H}_\phi}|\phi_{t+\epsilon}>\\
	&=& e^{-\epsilon {\cal H}_n(t)}\int Dp e^{i\int p(\phi_{t+\epsilon}-\phi_t)}
\end{eqnarray}
and yields, dropping the subscript $t$ in $\phi_t$, the Hamiltonian for the forward rates, namely
\begin{eqnarray}
\label{hphi}
{\cal H}_\phi (t)=-\frac{1}{2f_0^{2} }\int \sigma_0 e^{\nu\phi}(x) D(x,x';t)\sigma_0 e^{\nu\phi}(x')\frac{\delta^2}{\delta \phi(x)\delta \phi(x')}  
			-\frac{1}{f_0 }\int \alpha  \frac{\delta}{\delta \phi} \nonumber\\
\end{eqnarray}
The Hamiltonian is non-Hermetian with complex eigenvalues. Although this would problematic in physics, this is not so in finance since the Hamiltonian is not a physical quantity (such as energy) whose eigenvalues are observables, and hence is not required to have real eigenvalues.

\subsection{Hamiltonian for the Forward Rates and Volatility Quantum Fields}
We now consider the case when both the forward rates and its volatility fluctuate independently and are represented by separate quantum fields. We hence examine the Lagrangian given in eq.(\ref{stochasI}), namely
\begin{eqnarray}
\label{stochas2}
{\cal L}(t,x) &=& -\frac{1}{2(1-\rho^2)} \left(\frac{\frac{\partial f}{\partial t} - \alpha}{\sigma} - \rho \frac{\frac{\partial h}{\partial t} - \beta}{\xi}\right)^2 -
\frac{1}{2} \left(\frac{\frac{\partial h}{\partial t} -
\beta}{\xi}\right)^2 \nonumber\\
&& - \frac{1}{2\mu^2}\left(\frac{\partial}{\partial x}
\left(\frac{\frac{\partial f}{\partial t} - \alpha}{\sigma} \right)\right)^2
-\frac{1}{2 \kappa^2}
\left(\frac{\partial}{\partial x}\left(\frac{\frac{\partial
h}{\partial t} - \beta}{\xi}\right)\right)^2  
\end{eqnarray}
where recall
\begin{eqnarray}
-\infty \leq f(t,x),h(t,x) \leq + \infty
\end{eqnarray}

Discretizing time, and for notational simplicity suppressing the time and maturity labels, we write the Lagrangian ${\cal L}$ in matrix notation as follows
\begin{eqnarray}
\label{lag}
S(n)=-\frac{1}{2 \epsilon}\int
\left[ \begin{array}{ll} \sigma^{-1}A & \xi^{-1}B \end{array} \right](x)
{\cal M}(x,x';t)
\left[ \begin{array}{l}  \sigma^{-1}A \\
 \xi^{-1}B \end{array} \right](x')   
\end{eqnarray}
where
\begin{eqnarray}
\label{mtrx}
{\cal M}(x,x';t)=
\left[ \begin{array}{ll}  
 \frac{1}{1-\rho^2}-\frac{1}{\mu^2}\frac{\partial^2}{\partial x^2} & ~~~~- \frac{\rho}{1-\rho^2} \\
 ~- \frac{\rho}{1-\rho^2} & \frac{1}{1-\rho^2}-\frac{1}{\kappa^2}\frac{\partial^2}{\partial x^2}
\end{array} \right] \delta(x-x')
\end{eqnarray}
and
\begin{eqnarray}
A &\equiv& f_{t+\epsilon} - \tilde{f}_t - \epsilon \alpha \\
B &\equiv& h_{t+\epsilon} - \tilde{h}_t - \epsilon \beta 
\end{eqnarray}
Note that in obtaining eq.(\ref{lag}) for $S(n)$ we have used the boundary conditions on the fields given in eqs.(\ref{fbx}) and (\ref{hx}).

We rewrite eq.(\ref{lag}) using Gaussian integration and obtain (ignoring irrelevant constants)
\begin{eqnarray}
\label{pql}
&&e^{S(n)}\nonumber\\
&&=\prod_{x}\int dp(x)dq(x)
					\exp\big (-\frac{\epsilon}{2}\int \left[ \begin{array}{ll} p & q \end{array} \right]
	{\cal M}^{-1}\left[ \begin{array}{l}  p \\ q \end{array} \right] + i\int
	\left[ \begin{array}{ll} p & q \end{array} \right]\left[ \begin{array}{l}  \sigma^{-1}A \\ \xi^{-1}B  \end{array} \right] \big)
	\nonumber \\  
\end{eqnarray}

Recall from eq.(\ref{pim}) that the Feynman path integral has a non-trivial measure $\sigma^{-1}(t,x)$, and in obtaining the Hamiltonianwe we need to take this into account . 

Define the measure term by
\begin{eqnarray}
\rho_n \equiv\prod_{x}\sigma^{-1}(x)
\end{eqnarray} 
and rescale the $p$ and $q$ variables in eq.(\ref{pql}) for each $x$ as
\begin{eqnarray}
p&\rightarrow& \sigma p \\
q&\rightarrow& \xi q
\end{eqnarray}

We then obtain from eq.(\ref{pql}) that
\begin{eqnarray}
\rho_n e^{S(n)}&=&\int DpDq \times \nonumber\\
	\exp \big (-\frac{\epsilon}{2}\int  \left[ \begin{array}{ll} \sigma p & \xi q \end{array} \right]
	{\cal M}^{-1}\left[ \begin{array}{l}  \sigma p \\ \xi q \end{array} \right]&+&i\int 
\left[ \begin{array}{ll} p & q \end{array} \right]\left[ \begin{array}{l} f- \tilde{f}- \epsilon \alpha \\ h-\tilde{h}-\epsilon \beta  \end{array} \right] \big ) \nonumber\\
\end{eqnarray}
showing that the measure term cancels out. We hence have from above
\begin{eqnarray}  	
\rho_n e^{S(n)}
  &=&<\tilde{f};\tilde{h}|e^{-\epsilon {\cal H}_n}|f;h>\\
	&=& e^{-\epsilon {\cal H}_n(t)}\int DpDq e^{i\int p(f-\tilde{f})+i\int q(h-\tilde{h})}
\end{eqnarray}
and which yields the Hamiltonian for the forward rates and volatility as independent quantum fields given by 
\begin{eqnarray}
\label{hfh}
{\cal H}(t)=\frac{1}{2}\int \left[ \begin{array}{ll} \sigma \frac{\delta}{i\delta \tilde{f}} & \xi \frac{\delta}{i\delta \tilde{h}} \end{array} \right]
	{\cal M}^{-1}\left[ \begin{array}{l}  \sigma \frac{\delta}{i\delta \tilde{f}} \\ \xi \frac{\delta}{i\delta \tilde{h}} \end{array} \right] 
			-\int \{\alpha \frac{\delta}{\delta \tilde{f}}+ \beta \frac{\delta}{\delta \tilde{h}}\}
\end{eqnarray}
From eq.(\ref{mtrx}) we have
\begin{eqnarray}
{\cal M}^{-1}(x,x';t)= c
\left[ \begin{array}{ll}  
 D_--D_+ +\frac{1-\rho^2}{\kappa^2}(r_+D_+ - r_-D_-) & ~~~~~~~~~ \rho(D_--D_+) \\
  ~~~~~~~~~ \rho(D_--D_+) & D_--D_+ +\frac{1-\rho^2}{\mu^2}(r_+D_+ - r_-D_-)
\end{array} \right] \nonumber
\end{eqnarray}
where
\begin{eqnarray}
c&=&\frac{\mu^2 \kappa^2}{\sqrt{(\kappa^2-\mu^2)^2+4\rho^2 \mu^2 \kappa^2}}\\
r_{\pm}&=&\frac{1}{2(1-\rho^2)}[\mu^2+\kappa^2 \pm 
\sqrt{(\kappa^2-\mu^2)^2+4\rho^2 \mu^2 \kappa^2}~~]\\
(-\frac{\partial^2}{\partial x^2}&+&r_{\pm})D_{\pm}(x,x';t)=\delta(x-x')~~\mathrm{with~Neumann~boundary~conditions}\nonumber\\
\end{eqnarray}
For solving the no aribitrage condition, we will need
\begin{eqnarray}
G(x;x',t)&\equiv &{\cal M}^{-1}_{11}(x,x';t)\\
	&=&\frac{\mu^2}{\sqrt{(\kappa^2-\mu^2)^2+4\rho^2 \mu^2 \kappa^2}}
						\big [\kappa^2(D_--D_+) +(1-\rho^2)(r_+D_+ - r_-D_-)\big ]\nonumber\\
\end{eqnarray}

\section {Hamiltonian Formulation of No Arbitrage}
The principle of no arbitrage is central to the theory of finance, and a path integral formulation of this principle is given in \cite{beb1}. For the case of deterministic volatility, the Lagrangian for the forward rates as given in eq.(\ref{lf}) is quadratic, and hence the condition of no abitrage could be solved exactly by performing a Gaussian path integration \cite{beb1}. 

For the case of stochastic volatility, the Lagrangian is nonlinear and hence the condition of no arbitrage cannot be solved explicitly using the path integral; for this reason we reformulate the no arbitrage condition using the Hamiltonian. We will show that the Hamiltonian formulation even for the nonlinear theory of the forward rates with stochastic volatility allows for an exact solution of the no arbitrage condition.

We first derive the Hamiltonian formulation of no arbitrage for the case of a single security 
$S$, since the derivation  for the forward rates is more complex.

\subsection{No Arbitrage for a Single Security} 
Consider an option on a  security $S=e^x$ that  matures at time $T$ and has a payoff function given by $g(x,K)$, where $K$ is the strike price. As discussed in \cite{beb1}, the risk-free evolution of the  security is  given by the Hamilitonian ${\cal H}_s$, with the value of the option at time $t<T$ being given by 
\begin{eqnarray}
\label{secprob}
f(t,x) = e^{-r(T - t)}  \int_{-\infty}^{\infty}dx'
 <x|  e^{-(T - t){\cal H}_s}   |x'> g(x')
\end{eqnarray}
($r$ is a given constant risk-free spot interest rate). 

The martingale condition for the risk-free evolution of the security is that the discounted evolution of the future price of the security at some future time, say $t_*$, is equal, on the average, to the price of the security at earlier time $t$. The equation for the martingale condition states
\begin{eqnarray}
S(x(t))= E_{[t,t_*]}[e^{-(t_*-t)r}S(x(t_*))]
\end{eqnarray}
where the notation $E_{[t,t_*]}[Y]$ denotes  the average value of $Y$ over all the stochastic variables  in the time interval $(t,t_*]$. From eq.(\ref{secprob}), we have
\begin{eqnarray}
S(x) &=& e^{-r(t_* - t)}  \int_{-\infty}^{\infty}dx'
 <x|  e^{-(t_* - t){\cal H}_s}   |x'> S(x')\\
\label{secmart}
\Rightarrow <x|S>&=&\int_{-\infty}^{\infty}dx'
 <x|  e^{-(t_* - t)({\cal H}_s+r)}|x'> <x'|S>
\end{eqnarray}
Using the completeness equation for a single security given by
\begin{eqnarray}
{\cal I}=\int_{-\infty}^{\infty}dx |x><x|  
\end{eqnarray}
yields from eq.(\ref{secmart}), the   operator equation
\begin{eqnarray}
|S>=e^{-(t_* - t)({\cal H}_s+r)}|S>=0
\end{eqnarray}
Since time $t_*$ is arbitrary, we have
\begin{eqnarray}
\label{hsec}
 ({\cal H}_s+r)|S>=0
\end{eqnarray}
One can easily verify  that the Black-Scholes Hamiltonian for both the case of deterministic and stochastic volatility given in \cite{beb2} satisfies eq.(\ref{hsec}).

The result given in eq.(\ref{hsec}). shows that the existence of a martingale measure is equivalent to a risk-free Hamiltonian that annihilates (upto a constant $r$) the underlying security $S$. We will see that a similar condition holds for the Hamiltonian of the forward rates, but with a number of complications arising from the non-trivial domain of the forward rates and that the spot rate $r(t)=f(t,t)$ is itself a stochastic quantity.

\subsection{No Arbitrage for the Forward Rates}
The principle of no aribitrage states that the price of the bond $P(t_*,T)$ at some future time $T>t_*>t$ is the equal to the price of the bond at time $t$, discounted by the risk free interest rate $r(t)=f(t,t)$. In other words
\begin{eqnarray}
\label{m}
P(t,T)=E_{[t,t_*]}[e^{-\int_{t}^{t_*}r(t)dt} P(t_*,T)]
\end{eqnarray}
where, as before, $E_{[t,t_*]}[Y]$ denotes  the average value of $Y$ over all the stochastic variables  in the time interval $(t,t_*)$. 

\begin{figure}[h]
\begin{center}
\epsfig{file=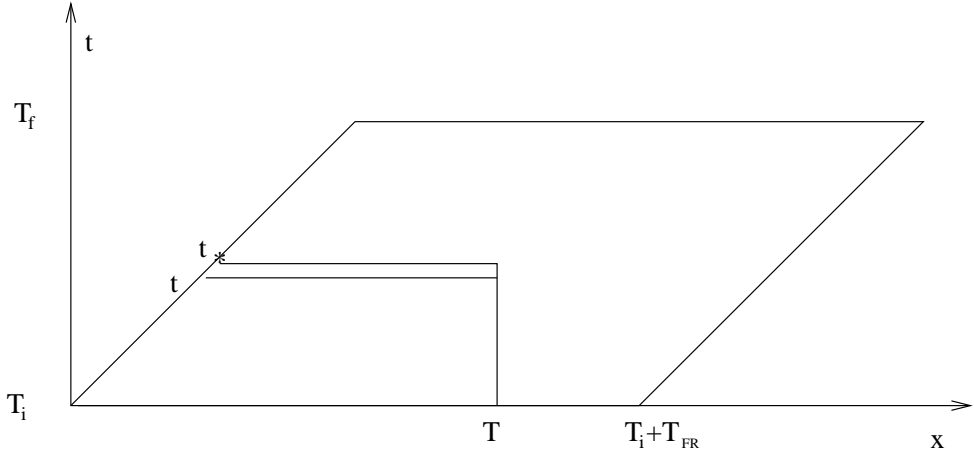,height=6cm}
\caption{Domains for No Arbitrage based on Treasury Bonds}
\label{tbnoarb}
\end{center}
\end{figure}

In terms of the Feynman path integral, eq(\ref{m}) yields (for measure $\rho$)
\begin{eqnarray}
\label{mfp}
P(t,T)=\int Df \rho[f] e^{-\int_{t}^{t_*}r(t)dt}e^{S[f]} P(t_*,T)
\end{eqnarray}

There are two domains involved involved in the path integral given above in eq.(\ref{mfp}), namely the domain for the Treasury bonds that is nested inside the domain of the forward rates. These domains are shown in Figure \ref{tbnoarb}.

Although written in an integral form, the condition eq.(\ref{mfp}) is clearly a differential condition since it holds for any value of $t_*$. Hence we take $t_*=t+\epsilon$. The reason that we need to consider only an infinitesimal change for the  forward rates is due to the time dependent nature of the state space ${\cal V}_t$. For an infinitesimal evolution in time, the funtional integral in eq.(\ref{mfp}) collapses to an integration over the final time variables $\tilde{f}_{t+\epsilon}$  on time slice $t+\epsilon$, that is
\begin{eqnarray}
\label{mfpe}
P(t,T)=\int D\tilde{f}_{t+\epsilon} \rho_{t+\epsilon}~e^{-\epsilon f(t,t)}
					e^{\epsilon\int{\cal L}[f,\tilde{f}]} P[\tilde{f};t+\epsilon,T]
\end{eqnarray}
We re-write above equation in the language of state vectors, namely that 
\begin{eqnarray}
\label{mfss}
<f_t|P(t,T)>=\int D\tilde{f}_{t+\epsilon} <f_t|e^{-\epsilon f(t,t)}
					e^{-\epsilon {\cal H}}|\tilde{f}_{t+\epsilon}>
														<\tilde{f}_{t+\epsilon}|P(t+\epsilon,T)>
\end{eqnarray}
We have, from the completeness equation given in eq.(\ref{comeq}), that
\begin{eqnarray}
\label{ce}
{\cal I}_{t+\epsilon}=\int D\tilde{f}_{t+\epsilon} |\tilde{f}_{t+\epsilon}><\tilde{f}_{t+\epsilon}|
\end{eqnarray}
Hence we have from eq.(\ref {mfss}) that
\begin{eqnarray}
\label{mhs}
<f_t|P(t,T)>&=& <f_t|e^{-\epsilon f(t,t)}e^{-\epsilon {\cal H}(t)}|P(t+\epsilon,T)>\\
\Rightarrow |P(t,T)>&=& e^{-\epsilon f(t,t)}e^{-\epsilon {\cal H}(t)}|P(t+\epsilon,T)>
\end{eqnarray}
It can be verified, using the explicit representation of the zero coupon 
bond given in eq.(\ref{zcb}), that 
\begin{eqnarray}
\label{minfis}
e^{+\epsilon f(t,t)}|P(t,T)>=|P(t+\epsilon,T)>
\end{eqnarray}
Hence, we have
\begin{eqnarray}
|P(t+\epsilon,T)>&=& e^{-\epsilon {\cal H}(t)}|P(t+\epsilon,T)>\\
\Rightarrow {\cal H}(t) |P(t+\epsilon,T)>&=&0
\end{eqnarray}
Since there was nothing special about the bond that we considered, we arrive at the differential formulation of no arbitrage, namely that any zero coupon bond -- and consequently any coupon bond -- is annihilated by the Hamiltonian ${\cal H}$. That is
\begin{eqnarray}
\label{hvac}
{\cal H}(t)|P(t,T)>=0~\mathrm{for~all~}t,~T
\end{eqnarray}
Note the similarity of above equation with the case of a single security given in 
eq.(\ref{hsec}). The role of the discounting factor is, however, very different in the two cases. The spot rate $r$ is a constant for the case of a security, whereas in the case of the forward rates -- given the difference in the domain for the state space at two different instances --  the discounting by the spot rate is precisely the factor required to transform the later time Treasury bond $P(t+\epsilon,T)$  to the one at an earlier time, namely $P(t,T)$. 

\section{No Arbitrage Condition for Stochastic Volatility}
Armed with the Hamiltonian for forward rates with stochastic volatility given in eqs.(\ref{hphi}) and (\ref{hfh}), we apply the no arbitrage conditon obtained in eq.(\ref{hvac}), namely
\begin{eqnarray}
\label{hamno}
{\cal H}(t)|P(t,T)>=0
\end{eqnarray}
or more explicity
\begin{eqnarray}
\label{hnoarb2}
<f_t|{\cal H}(t)|P(t,T)>={\cal H}(t)e^{-\int_t^T dx f(t,x)}=0
\end{eqnarray}

\subsection{No Arbitrage for Volatililty a Function of Forward Rates}
Recall the zero coupon bond is given by
\begin{eqnarray}
P(t,T)=\exp\big(-f_0\int_t^T dx e^{\phi(t,x)}\big) 
\end{eqnarray}
and which yields
\begin{eqnarray}
\label{fdiff}
\frac{\delta}{\delta \phi(t,x)}P(t,T) & =  & \left\{ 
\begin{array}{ll}   
-f_0 e^{\phi(t,x)}P(t,T), & t<x<T \nonumber\\
0, & x>T \end{array} \right.\\  
\end{eqnarray}
We have from eqs.(\ref{hnoarb2}) and (\ref{hphi}) that
\begin{eqnarray}
\Big [-\frac{1}{2}\int_t^Tdxdx' \sigma_0e^{\nu \phi(t,x)}f(t,x)D(x,x';t)\sigma_0 e^{\nu \phi(t,x')}f(t,x') \nonumber\\
    +\int_t^T dx \alpha(t,x)f(t,x)\Big ]P(t,T)=0 \nonumber\\
\label{al}
\Rightarrow \alpha(t,x)=\frac{\sigma_0}{f_0}e^{\nu \phi(t,x)}\int_t^x~dx'D(x,x';t)\sigma_0e^{\nu \phi(t,x')}f(t,x')&& \\
				\mathrm{:~Condition~of~No~Arbitrage}\nonumber
\end{eqnarray}

Note that the no arbitrage condition given above is {\bf not} contained in the HJM-class of solutions for the drift velocity which are all quadratic in the volatility fields \cite{an}; the appearance of the forward rates $f(t,x)$ directly in the drift velocity emerges naturally in the field theoretic formulation, and is a reflection of the kinetic term in the Lagrangian  for the case of $f\in [-\infty,+\infty]$ -- namely $\displaystyle (\frac{\partial f}{\partial t})^2$   -- being replaced by $\displaystyle (\frac{\partial \phi}{\partial t})^2$ for $f\in [0,+\infty]$. 

We can in fact prove a more general result for the action $S[\phi]$ for the case of when stochastic volatility is a function of the forward rates. We write the most general Lagrangian as
\begin{eqnarray}
{\cal L}_{\mathrm{general}}&=&{\cal L}[\phi]+\int U(t,x)\frac{\partial \phi}{\partial t}+\int W(t,x)\\
				U,W &:& \mathrm{arbitrary~local~functions~of~}f(t,x)
\end{eqnarray}
The no aribitrage condition then yields that
\begin{eqnarray}
U(t,x)=W(t,x)=0
\end{eqnarray}
In particular, a string tension term in the Lagrangian has the form 
\begin{eqnarray}
W(t,x)\propto (\frac{\partial f}{\partial x})^2
\end{eqnarray}
and is forbidden by the condition of no arbitrage.

\subsection{No Arbitrage for Volatililty as an Independent Quantum Field}
From the Hamiltonian given in eq.(\ref{hfh}) we see that, as in the case above, $\displaystyle \frac{\delta}{\delta h}$ yields zero in the eq. (\ref{hnoarb2}) since the zero coupon bond does not depend explicitly on the volatility field. Using the fact that 
\begin{eqnarray}
\label{fdiff}
\frac{\delta^m}{\delta f^m(t,x)}~ e^{-\int_t^T dx f(t,x)} & =  & \left\{ 
\begin{array}{ll}   
(-1)^m~e^{-\int_t^T dx f(t,x)}, & t<x<T \nonumber\\
0, & x>T \end{array} \right.\\  
\end{eqnarray}
We hence have from eqs.(\ref{hnoarb2}), (\ref{hfh}) and (\ref{fdiff})
\begin{eqnarray}
\Big [-\frac{1}{2}\int_t^Tdxdx' \sigma(t,x)G(x,x';t)\sigma(t,x')
 +\int_t^T dx \alpha(t,x)\Big ]P(t,T)=0 \nonumber\\
\label{al}
\Rightarrow \alpha(t,x)=\sigma(t,x)\int_t^x~dx'G(x,x';t)\sigma(t,x') \\
				\mathrm{:~Condition~of~No~Arbitrage}\nonumber
\end{eqnarray}

Since  there is no instrument in the financial markets at present that trades in volatility of  the forward rates, we cannot apply the condition of martingale to the volatilty field, and in particular, we cannot obtain the drift velocity of the volatility field, namely $\beta(t,x)$, to be a function of the other fields and parameters of the theory. For this reason $\beta$  has to be determined empirically from the market.
To obtain the limit of volatility being determistic,  we need to take the limit of $\xi, \rho$ and $\kappa \rightarrow 0$. We then have
\begin{eqnarray}
\xi, \rho, \kappa &\rightarrow& 0  \\
r_+ &\rightarrow& \mu \\
r_- &\rightarrow& 0 \\
G(x,x';t) &\rightarrow&  D(x,x';t) 
\end{eqnarray}
with propagator $D(x,x';t)$ given by eq.(\ref{prop}).

By a remarkable set of identities,  it can be shown that the propagator $D(x,x';t)$ given in eq.(\ref{prop}) above is  exactly equal to the one obtained in ref \cite{beb1} using path integral techniques. Hence the no arbitrage condition obtained for $\alpha$ for the case of  deterministic volatility using the Hamiltonian condition is the same as the one obtained earlier path integration.

Incorporating  the expression for $\alpha(t,x)$ given in eq.(\ref{al}) into the Lagrangian yields the final result. For notational convenience define the following non-local function of the volatility field
\begin{eqnarray}
v(t,x)=\int_t^x~dx'G(x,x';t)\sigma(t,x')
\end{eqnarray}

We hence obtain
\begin{eqnarray}
\label{lnoarb}
{\cal L}(t,x) &=& -\frac{1}{2(1-\rho^2)} \left(\sigma^{-1}\frac{\partial f}{\partial t} - v - \rho \frac{\frac{\partial h}{\partial t} - \beta}{\xi}\right)^2 -
\frac{1}{2} \left(\frac{\frac{\partial h}{\partial t} -
\beta}{\xi}\right)^2 \nonumber\\
&& - \frac{1}{2\mu^2}\left(\frac{\partial}{\partial x}
\left(\sigma^{-1}\frac{\partial f}{\partial t} - v \right)\right)^2
-\frac{1}{2 \kappa^2}
\left(\frac{\partial}{\partial x}\left(\frac{\frac{\partial
h}{\partial t} - \beta}{\xi}\right)\right)^2  
\end{eqnarray}
The Lagrangian given by the equation above is a complete description of the theory of forward rates with stochastic volatility. All the parameters in the theory, namely the function $\beta(t,x)$ and the parameters $\mu, \kappa, \xi$ and $\rho$  need to be determined from market data. Due to the presence of the field $v(t,x)$ we find that the Lagrangian is non-local, with the function $G(x,x';t)$ containing all the information regarding the absence of arbitrage.

There are two further generalizations that we can make of the Lagrangian obtained in eq.(\ref{lnoarb}), namely that the forward rate can be made positive, that is, $f>0$, and that the propagator $G(x,x';t)$ can include more complex effects arising from a  dependence on maturity of the rigidity parameter $\mu$.
 
\section {Conclusions}
We made a generalization of the field theory model for the forward rates to account for stochastic volatility by treating volatility either as a function of the forward rates or as an independent quantum field. In both cases, the Feynman path integral could be naturally extended to account for stochastic volatility. 

For the case of deterministic volatility, it was found in \cite{beb1} that in effect the two dimensional quantum field theory reduced to a one-dimensional problem due to the specific nature of the Lagrangian. However, on treating volatility as a quantum field, the theory is now irreducibly two-dimensional, and displays all the features of a quantum field theory.

To exactly solve for the no arbitrage condition for stochastic volatility, we had to recast the condition of no arbitrage as a condition involving the Hamiltonian of the system. To obtain the Hamiltonian of the forward rates, we were in turn led to an analysis of the underlying state space of the system, which turned out to be non-trivial due to the parallelogram domain on which the forward rates are defined. The Hamiltonian for the forward rates is an independent formulation of the theory of the forward rates, and can lead to new insights on the behaviour of the forward rates.

The model for the forward rates with stochastic volatility has a number of free parameters that can only be determined by studying the market. Hence on needs to be numerically analyze the model so as to calibrate it, and to test its ability to explain the market's behaviour. The first step in this direction has been taken in \cite{b2} and these calculations are now being extended to the case of stochastic volatility.

\section*{Acknowledgments}
I have benefitted from many conversations with Srikant Marakani. I have would also like to thank Mitch Warachka, Rajesh Parwani and Omar Foda for valuable discussions. I thank Jean-Phillipe Bouchaud and Science and Finance for the data on the Eurodollar futures.

\section*{Appendix}
We explicitly evaluate the propagator $D(x,x';t)$. 
Note that the normalized eigenfunctions on the interval $[t,t+T_{FR}]$ that satisfy the Neumann condition of vanishing derivatives at $x=t$ and $x=t+T_{FR}$ are given by
\begin{eqnarray}
\psi_m(x)&=&\frac{1}{\sqrt{T_{FR}}}\cos\{\frac{m\pi (x-t)}{T_{FR}}\}; m=0,1,2,3, ..... \infty 
\end{eqnarray}
which satisfy the eigenvalue equation
\begin{eqnarray}
(-\frac{1}{\mu^2}\frac{\partial^2}{\partial x^2}+1)\psi_m(x)&=& \big [(\frac{m\pi}{\mu T_{FR}})^2+1 \big ] \psi_m(x)
\end{eqnarray}
Hence we have
\begin{eqnarray}
D(x,x';t)&=&\frac{1}{2 T_{FR}}\psi^2_0(x)+\frac{1}{T_{FR}} \sum_{m=1}^{\infty}  \frac{\psi_m(x)\psi_m(x')}{(\frac{m\pi}{\mu T_{FR}})^2+1 }
\end{eqnarray}
and which yields, after some simplifications
\begin{eqnarray}
D(x,x';t)=\frac{\mu}{2 \sinh(\mu T_{FR})} 
			 [\cosh(\mu T_{FR}&-&\mu |x-x'|) \nonumber\\
				+ \cosh(\mu T_{FR}&-&\mu (x+x'-2t))]
\end{eqnarray}

\end{document}